\documentclass[12pt]{article}
\newcommand{\nc}{\newcommand}
\nc{\bib}{\bibitem}
\nc{\al}{\alpha}
\nc{\g}{\gamma}
\nc{\G}{\Gamma}
\nc{\D}{\Delta}
\nc{\eps}{\epsilon}
\nc{\la}{\lambda}
\nc{\La}{\Lambda}
\nc{\var}{\varphi}
\nc{\cg}{{\cal G}}
\nc{\pa}{\partial}
\nc{\nn}{\nonumber \\ }
\nc{\hf}{\frac{1}{2}}  
\nc{\dz}{\frac{dz}{2\pi i}}
\nc{\bin}[2]{\left (\begin{array}{c} {#1}\\ {#2} \end{array}\right )}
\nc{\ben}{\begin{equation}}
\nc{\een}{\end{equation}}
\nc{\bea}{\begin{eqnarray}}
\nc{\eea}{\end{eqnarray}}
\nc{\bra}[1]{\langle {#1}|}
\nc{\ket}[1]{|{#1}\rangle}
\nc{\C}{\mbox{\hspace{1.24mm}\rule{0.2mm}{2.5mm}\hspace{-2.7mm} C}}
\nc{\Nat}{\mbox{\hspace{.04mm}\rule{0.2mm}{2.8mm}\hspace{-1.5mm} N}}
\newcommand{\R}{\mbox{\hspace{.04mm}\rule{0.2mm}{2.8mm}\hspace{-1.5mm} R}}
\nc{\HH}{\mbox{\hspace{.04mm}\rule{0.2mm}{2.8mm}\hspace{-1.5mm} H}}

\def\vvdots{\mathinner{\mkern1mu\raise1pt\vbox{\kern7pt\hbox{.}}\mkern2mu
 \raise4pt\hbox{.}\mkern2mu\raise7pt\hbox{.}\mkern1mu}}

\begin{document}

\topmargin -5mm
\oddsidemargin 5mm

\begin{titlepage}
\setcounter{page}{0}

\vspace{8mm}
\begin{center}
{\Large {\bf On string backgrounds and (logarithmic) CFT\footnote{Based
on invited talk presented at the IX-th Workshop on Mathematical Physics 
and Applications, Rabat, 23-25 February 2004.}}}

\vspace{15mm}
{\Large J{\o}rgen Rasmussen}\\[.3cm] 
{\em Centre de recherches math\'ematiques, Universit\'e de Montr\'eal}\\ 
{\em Case postale 6128, 
succursale centre-ville, Montr\'eal, Qc, Canada H3C 3J7}\\[.3cm]
rasmusse@crm.umontreal.ca

\end{center}

\vspace{10mm}
\centerline{{\bf{Abstract}}}
\vskip.4cm
\noindent
We discuss the link between string backgrounds
and the associated world-sheet CFTs. In the search for new
backgrounds and CFTs, Penrose limits and Lie algebra
contractions are important tools. 
The Nappi-Witten construction and the recently discovered
logarithmic CFT by Bakas and Sfetsos, are
considered as illustrations. We also speculate on possible extensions.
\end{titlepage}
\newpage
\renewcommand{\thefootnote}{\arabic{footnote}}
\setcounter{footnote}{0}

\section{Introduction}

A common problem in string theory is to understand the links between
the world-sheet description and the space-time (or target-space)
physics. A main focus here will be on a particular aspect of this,
namely the study of which string backgrounds may be associated
to which world-sheet conformal field theories (CFTs).
As will be discussed, an intriguing observation is that
logarithmic CFT seems to enter the game \cite{BS}.
We refer to \cite{Flo,Gab} for recent reviews on logarithmic CFT.

Often the link is made explicit by considering the string theory as a
(non-linear) $\sigma$-model with a given background metric.
It is then identified with some sort of Wess-Zumino-Witten (WZW)
model on the world sheet. Depending on the WZW model being
based on a (non-)semi-simple group, (non-)compact or perhaps
a coset in terms of a gauged WZW action, we will get different
string backgrounds. Work related to this may be found in
\cite{NW,KK,Sfe010,ORS,Moh,FS,KM,AO,ST,BS,Sfe} 
and references therein. It turns out that some of the more
exotic constructions on the world sheet can be obtained by
considering so-called Lie algebra contractions of the Lie 
algebras underlying some more conventional initial WZW models.
Lie algebra contractions are sometimes referred to as
In\"on\"u-Wigner or Saletan contractions. One could also consider
more general CFTs where the Lie algebra contractions 
are replaced by linear, but singular, maps of the set
of primary fields. The new model (which may not even be
an ordinary CFT) is obtained by considering the singular
limit. We suggest to refer to these more general constructions
as operator product algebra (OPA) contractions.

On the space-time side, we can obtain new geometries
by considering Penrose limits of existing geometries.
As a result, one obtains pp-wave backgrounds which --in a
sense-- differ only minimally from flat backgrounds.
The reason is that the curvature effects are rendered controllable
by the existence of a covariantly constant null Killing vector.
We refer to \cite{SS} for a recent review on plane waves and
their applications.

The aim of this talk is to indicate how Lie algebra or OPA contractions
and Penrose limits may be seen as going hand in hand, by 
presenting a couple of
examples based on \cite{NW,BS} and some speculations. As a
recent offspring of their affair,
we recognize the emergence of a logarithmic CFT \cite{BS}.

\section{String backgrounds and CFT}

To get an impression of how the link between the world sheet
and string background is established, let us consider
the ordinary WZW model based on an action like
\ben
 S\ =\ \frac{1}{4\pi}\int_\Sigma(g^{-1}dg)^2\ 
  +\ \frac{i}{6\pi}\int_B(g^{-1}dg)^3
\label{WZW}
\een
where $B$ is a three-dimensional space with boundary
given by the two-dimensional surface $\Sigma$. $g$ takes
values in a Lie group, so the one-forms can be expressed in terms of
the Lie algebra generators:
\ben
 g^{-1}\pa_\al g\ =\ A_\al^aJ_a\ ,\ \ \ \ \ \ \
  \left[J_a,J_b\right]\ =\ {f_{ab}}^cJ_c
\label{A}
\een
For the model to be well-defined, we need a bilinear and symmetric
two-form, $\Omega$, on the Lie algebra satisfying
\ben
 {f_{ab}}^d\Omega_{cd}+{f_{ac}}^d\Omega_{bd}\ =\ 0
\label{omega}
\een
This is referred to as invariance and corresponds to imposing
the Jacobi identities on the affine extension of the Lie
algebra. Furthermore, the two-form must be non-degenerate, that is,
it must be invertible. With all this satisfied, we can rewrite
the action (\ref{WZW}) as
\ben
 S\ =\ \frac{1}{4\pi}\int_\Sigma d^2x\Omega_{ab}A^a_\al A^{b\al}\ 
  +\ \frac{i}{12\pi}\int_B
   d^3x\eps^{\al\beta\g}A^a_\al A^b_\beta A^c_\g\Omega_{cd}{f_{ab}}^d
\label{WZWA}
\een

The two-form $\Omega$ trivially exists when $G$ is semi-simple and is then
given by the Cartan-Killing form, the trace in the adjoint representation.
For non-semi-simple groups, this form is degenerate. Nappi 
and Witten provided an example in \cite{NW} of a 
WZW model based on a non-semi-simple group admitting a 
non-degenerate two-form. We shall return to it below.

Now, choosing a parametrization of the group elements as
\ben
 g\ =\ e^{j^aJ_a}...e^{j^bJ_b}
\label{g}
\een
one can write the WZW action in terms of the coefficients $j^a$ and
their derivatives. This is then compared to the $\sigma$-model
\ben
 S_\sigma\ =\ \int\left(G_{MN}\pa_\al X^M\pa^\al X^N
  +iB_{MN}\eps_{\al\beta}\pa^\al X^M\pa^\beta X^N\right)
\label{sigma}
\een
where $G_{MN}$ gives the space-time background metric, whereas
$B_{MN}$ is the anti-commuting tensor field. Thus, considering
the string theory as described by such a $\sigma$-model, this provides
the link between the world-sheet CFT and the string background.
Notice that the dimension
of the background is given by the dimension of the Lie group.

Different representations of the group elements (\ref{g}) will
result in different metrics, so the Campbell-Baker-Hausdorff
formula is seen to generate coordinate transformations on
the background.

An important result of Nappi and Witten was that a sufficiently funky choice
of WZW model could lead to a plane-wave background. 
In \cite{NW} they considered a central extension of the
two-dimensional Poincare or Euclidean algebra $E_2^c$:
\ben
 [J,P_i]=\eps_{ij}P_j\ ,\ \ \ \ [P_i,P_j]=\eps_{ij}T\ ,\ \ \ \ [T,J]=[T,P_i]=0
\label{E}
\een
Here $J$ generates rotations while $P_i$ generate translations.
$T$ is a central element and governs the extension.
In the order $(P_1,P_2,J,T)$, the most general two-form reads
\ben
 \Omega\ =\ k\left(\begin{array}{llll} 1&0&0&0\\ 0&1&0&0\\ 0&0&b&1\\
   0&0&1&0 \end{array}\right)
\label{b}
\een
and is of Lorentz signature $(+++-)$.
With the representation
\ben
 g\ =\ e^{x_1P_1}e^{uJ}e^{x_2P_1+vT}
\een
the associated metric is worked out to be
\ben
 \frac{1}{k}ds^2\ =\ dx_1^2+dx_2^2+2\cos(u)dx_1dx_2+2dudv+bdu^2
\label{dsNW}
\een
This is recognized as the metric of a plane wave.

One can now address the conformal invariance of the model
by showing that the one-loop $\beta$-function vanishes, fixing
the central charge to 4. One way of checking this non-perturbatively
is to consider the generalized Sugawara construction by evaluating
the so-called Virasoro master equation.
A more general argument for this background to be a good
choice in string theory is due to Horowitz and Steif \cite{HS}, 
who found that a broad class of pp-waves, being solutions to the
supergravity equations, do not receive $\al'$ corrections.

Here, instead, we turn to the affair of Penrose limits and
Lie algebra contractions.
Let us consider the WZW model based on $SU(2)\times\R$
where the second factor is generated by a time-like coordinate,
$y$. In this case one can write the metric as
\ben
 \frac{2}{k'}ds^2\ =\ d\theta_L^2+d\theta_R^2+d\phi^2
  +2\cos(\phi)d\theta_Ld\theta_R-dy^2
\label{dssu2}
\een
where $\theta$ and $\phi$ are angles parametrizing $SU(2)$.
For $\eps\neq0$, the coordinate transformation
\ben
 k'=2k/\eps\ ,\ \ \ \ \theta_L=\sqrt{\eps}x_1\ ,\ \ \ \ 
  \theta_R=\sqrt{\eps}x_2\ ,\ \ \ \ \phi=\eps v+u\ ,\ \ \ \ 
   y=(1-\eps b/2)u
\label{k'}
\een
is merely a linear transformation with a singularity at $\eps=0$.
However, if we consider the correlated limit where
\ben
 \eps\rightarrow0\ ,\ \ \ \ \ \ \ \ 2k=k'\eps\ \ \mbox{fixed}
\label{PL}
\een
the geometry changes and we end up with the Nappi-Witten (NW)
background (\ref{dsNW}). 
A correlated limit like (\ref{k'}) and (\ref{PL}) is called a Penrose limit.

This construction has an algebraic analogue on the world sheet.
To see this, let us consider the algebra $su(2)\oplus u(1)$ with
generators normalized as
\ben
  [J_x,J_y]=J_z\ ,\ \ \ \ [J_y,J_z]=J_x\ ,\ \ \ \ [J_z,J_x]=J_y\ ,\ \ \ \ 
   [U,J_\ast]=0
\label{su2u1}
\een
After the following change of basis
\ben
 \left(\begin{array}{l}P_1\\ P_2\\ J\\ T\end{array}\right)\ =\ 
  \left(\begin{array}{llll} a&0&0&0\\ 0&a&0&0\\ 0&0&1&\frac{-b}{2a}\\
   0&0&0&b \end{array}\right)
   \left(\begin{array}{l}J_x\\ J_y\\ J_z\\ U\end{array}\right)
\label{basis}
\een
one can easily write down the commutators of the new generators.
Even though the matrix is singular in the limit $a\rightarrow0$,
the resulting algebra nevertheless makes sense. This procedure
is an example of a Lie algebra contraction.
One finds that the new algebra is $E_2^c$, the algebra underlying
the NW construction.

We have thus illustrated the following schematic relation
between string backgrounds and world-sheet CFT, governed
by the interpretation of the string theory as a $\sigma$-model:\\[1cm]
\begin{tabular}{ccc}
  & ($\sigma$-model interpretation) &\\
 \underline{SPACETIME} & \hspace{.8cm} $<---------->$ \hspace{.8cm}
  &  \underline{WORLD SHEET}\\[.7cm]
  geometry && CFT (WZW model)\\
 $|$ && $|$\\
  $|$ && $|$\\
 $\begin{array}{c} correlated\\ (Penrose)\ limit \end{array}$  
  && $\begin{array}{c} Lie\ alg.\\ contraction \end{array}$\\
  $|$ && $|$\\
  $\downarrow$ && $\downarrow$\\
 geometry' & $<---------->$  & CFT'\\
 (NW plane-wave) & ($\sigma$-model interpretation)  
  & (non-semi-simple WZW)
\end{tabular}\\[1cm]
The two mediators are correlated (Penrose) limits and Lie algebra
contractions, respectively.

\section{String backgrounds and logarithmic CFT}

Another and more recent example illustrating the general
picture above is due to Bakas and Sfetsos \cite{BS}.
Extensions are discussed in \cite{Sfe,BJR}. 
It is based on a parafermionic model $SU(2)_N/U(1)_N$
times a time-like boson generating $U(1)_{-N}$.
Its metric can be written as
\ben
 \frac{1}{N}ds^2\ =\ -dt^2+d\theta^2+\cot^2(\theta)d\phi^2
\label{N}
\een
where $t$ represents the time-like coordinate.
The remaining part stems from an Euler-angle representation
of an $SU(2)$ WZW model gauged by a $U(1)$ subgroup.
The correlated limit of our interest is based on the transformation
\ben
 \theta=\eps v+u\ ,\ \ \ \ t=u\ ,\ \ \ \ \phi=\sqrt{\eps}x\ ,\ \ \ \ 
   N\eps=1
\label{eps}
\een
In the limit $\eps\rightarrow0$ the metric becomes
\ben
 ds^2\ =\ 2dudv+\cot^2(u)dx^2
\een
describing a plane-wave background.

On the world sheet, we start with a level-$N$
parafermionic CFT \cite{ZF} based on the coset $SU(2)_N/U(1)_N$.
It is generated by the two basic fields $\psi_1$ and
$\psi_1^\dagger$. Their conformal weights and the 
central charge are
\ben
 \D(\psi_1^{(\dagger)})=1-\frac{1}{N}\ ,\ \ \ \ \ \ c=\frac{2(N-1)}{N+2}
\label{Dc}
\een
We shall not go into details of the structure of the
operator product algebra nor the computation of
correlators. In the naive limit with $N$ approaching
infinity, the fields become bosons of dimension 1 and
the central charge is 2. Combined with an extra
$U(1)$ factor as above, we would then have three bosons
and $c=3$.

This parafermionic model has a classical counterpart \cite{BCR}
described by the target-space coordinates appearing in (\ref{N}).
In the classical model the transformation (\ref{eps}) corresponds to 
a field transformation. Motivated by this observation,
Bakas and Sfetsos considered the linear map
\ben
 \Phi=\eps\left(\frac{\sqrt{N}}{2}(\psi_1+\psi_1^\dagger)-J\right)
  ,\ \ \ \ \Psi=\frac{\sqrt{N}}{2}(\psi_1+\psi_1^\dagger)+J\ ,\ \ \ \ 
 P=\sqrt{\eps}\frac{\sqrt{N}}{2i}(\psi_1-\psi_1^\dagger)
\label{Phi}
\een
where $J$ is the $U(1)$ field.
All three fields in (\ref{Phi}) are seen to be self-conjugate.
It is emphasized that $\Phi$ and $\Psi$ do not have
well-defined weights for finite $N$ -- only in the limit 
$N\rightarrow\infty$.
This is important as it opens up for the possibility that,
even in the limit, these fields may not be primary fields after all.
Indeed, in the limit one finds that 
$\Psi$ is a logarithmic partner to the primary field $\Phi$:
\bea
 T(z)\Phi(w)&=&\frac{\Phi(w)}{(z-w)^2}
    +\frac{\pa_w\Phi(w)}{z-w}+{\cal O}(1)\nn
 T(z)\Psi(w)&=&\frac{\Psi(w)-\frac{1}{2}\Phi(w)}{(z-w)^2}
    +\frac{\pa_w\Psi(w)}{z-w}+{\cal O}(1) 
\label{T}
\eea
This is a basic feature of a logarithmic CFT where the
Virasoro generator $T$ no longer acts diagonally.
A canonical situation is illustrated by the (rank-two)
Jordan cell
\ben
 L_0\ket{\psi}=\D\ket{\psi}+\ket{\phi}\ ,\ \ \ \ L_0\ket{\phi}
   =\D\ket{\phi}
\label{Jordan}
\een
A CFT with such a construction is known to lead to
logarithmic dependencies of correlators, hence the term
{\em logarithmic} CFT.

The resulting logarithmic CFT in \cite{BS} has central charge 3
and is a new logarithmic CFT. This,
of course, is interesting in itself. It also points in the
direction of constructing other new models as we shall 
indicate in the following. Potentially even more
important, their work may turn out to provide an
application of logarithmic CFT, something we don't
have too many of.

Extensions of the work \cite{BS} would naturally be built
on cosets like
\ben
 (G_N/H_N)\ \times\ U(1)_{-N}
\label{GHU}
\een
where the $U(1)$ factor gives the time coordinate
ensuring a Lorentz signature of the target space.
Examples are provided in \cite{Sfe} and are based
on ordinary Lie groups. We have recently realized
that this seems to extend to supergroups as well.
The simplest set-up is based on the graded
parafermions:
\ben
 (OSp(1|2)_N/U(1)_N)\ \times\ U(1)_{-N}
\label{osp}
\een
and is discussed in \cite{BJR}.
The graded parafermionic CFT \cite{CRS} is generated by the
four basic fields $\psi_{1/2}$, $\psi_{1/2}^\dagger$,
$\psi_{1}$ and $\psi_{1}^\dagger$, the former two being odd.
Their dimensions and the central charge are given by
\ben
 \D(\psi_{1/2}^{(\dagger)})=1-\frac{1}{4N}\ ,\ \ \ \ 
 \D(\psi_1^{(\dagger)})=1-\frac{1}{N}\ ,\ \ \ \ c=\frac{-3}{2N+3}
\label{gradedD}
\een 
The construction mimics the bosonic scenario above,
and using conjugacy as a guiding principle,
there is hardly any freedom when generalizing (\ref{Phi}).
We thus suggest to supplement (\ref{Phi}) by the additional linear map
\ben
 \Upsilon^+=\frac{N^{1/4}}{\sqrt{2}}
  \left(\psi_{1/2}+\psi_{1/2}^\dagger\right)\ ,\ \ \ \ 
   \Upsilon^-=\frac{N^{-1/4}}{\sqrt{2}i}
    \left(\psi_{1/2}-\psi_{1/2}^\dagger\right)
\label{upsilon}
\een
The operator product algebra and the structure of correlators
in the logarithmic CFT emerging in the limit $N\rightarrow\infty$,
are discussed in \cite{BJR}.
This logarithmic CFT is generated by five spin-one fields, based
on (\ref{Phi}) and (\ref{upsilon}), two of which are odd,
and the central charge is 1.

\section{Speculations}

There are many interesting problems to look at in
connection with this subject, and the literature already 
contains numerous results. Here we list a few speculations with only
scarce reference to the existing literature.
\\[.3cm]
\noindent$\bullet$ It would be interesting to see how
far one could stretch the relationship between Lie
algebra or OPA contractions and correlated limits of linear, 
but singular, coordinate 
transformations\footnote{Before performing this transformation,
one may wish to 'prepare' by performing an ordinary and
invertible, but not necessarily linear, coordinate transformation.
This could be advantageous and would not change the geometry.}.
They are both formally of the form
\ben
 \left(\begin{array}{c}\mbox{}\\ {\rm new}\\ {\rm set}\\
  \mbox{}\end{array}\right)\ =\ \left(\begin{array}{c}\mbox{}\\
 {\rm singular}\\ {\rm matrix}\\ \mbox{}\end{array}\right)
 \left(\begin{array}{c}\mbox{}\\ {\rm old}\\ {\rm set}\\
  \mbox{}\end{array}\right)
\een
mapping the old set of generators (or fields) or coordinates into
the new set. Since the number of generators\footnote{ 
In the case of a world-sheet WZW model based on an
ordinary Lie group, this number equals the dimension 
of the Lie group, as already mentioned.}
is equal to the dimension of the geometry, one
could compare 'directly' the two matrices.
\\[.3cm]
\noindent$\bullet$ Another problem is to understand how
brane configurations change under the action of taking Penrose limits.
Some progress in this direction has been made in \cite{SF}, for example.
In this realm, one could also wonder about the fate of
duality under Penrose limits where correlated limits of
two dual descriptions would be considered.
\\[.3cm]
\noindent$\bullet$ The final point addressed here concerns a possible
'supersymmetric extension' of the NW construction. 
The basic observation is that the so-called non-reductive
$N=4$ superconformal algebra (SCA) found in \cite{non-red}
contains (an affine extension of) $E_2^c$ as a bosonic
subalgebra\footnote{The full structure of the SCA fixes $b=1$
in the two-form (\ref{omega}) of its subalgebra, 
and with $k$ playing the role of the level of
the affine extension, it is given in terms of the
central charge of the SCA: $k=\frac{c_{SCA}}{12}$.}. 
This SCA was constructed
as a non-trivial Lie algebra contraction of the well-known
large $N=4$ SCA \cite{STV}, further supporting the idea
that it may be of relevance in the present context.
It is also linked to the AdS/CFT correspondence as the
non-reductive $N=4$ SCA also contains the so-called
asymmetric $N=4$ SCA \cite{AdS/CFT} as a subalgebra.
This asymmetric $N=4$ SCA was constructed as a superconformal
extension of the Virasoro algebra generating the conformal 
transformations on the boundary of AdS$_3$.
\vskip.5cm
\noindent{\em Acknowledgements}
\vskip.1cm
\noindent The author thanks A. Belhaj and E.H. Saidi for the
invitation and very kind hospitality extended to him during
the workshop where this talk was presented. 
He also thanks P. Jacob for discussions.

\end{document}